\documentstyle[psfig]{l-aa}

\begin{document}

\thesaurus{03
           (11.01.2;
            11.09.1;
            11.19.3)}

\title{A search for molecular gas in high redshift radio galaxies}

\author{R.~van Ojik\inst{1}\and 
        H.J.A.~R\"ottgering\inst{1,2,3}\and
        P.P.~van der Werf\inst{1}\and 
        G.K.~Miley\inst{1}\and
        C.L.~Carilli\inst{1,4}\and 
        A.~Visser\inst{1}\and 
        K.G.~Isaak\inst{2}\and 
        M.~Lacy\inst{5}\and
        T.~Jenness\inst{2}\and 
        J.~Sleath\inst{2}\and 
        J.~Wink\inst{6}}

\offprints{H.J.A.~R\"ottgering (Leiden address)}

\institute{Leiden Observatory, P.O.~Box 9513, NL - 2300 RA Leiden,
The Netherlands \and
Mullard Radio Astronomy Observatory, Cavendish Laboratory,
Cambridge CB3 0HE, England \and 
Institute of Astronomy, Madingley Road, Cambridge CB3 0HA, England \and
NRAO, PO Box 0, Socorro, NM, 87801 \and 
Astrophysics, Department of Physics, Keble Road, Oxford, OX1 3RH, England \and
Institut de Radioastronomie Millim\'etrique, 300 Rue de la Piscine, 
F - 38406 St.~Martin d'H\`eres Cedex, France
}

\date{Accepted 27 September 1996}

\maketitle

\markboth{R.~van Ojik et al.: Molecular gas in high-$z$ radio galaxies}{} 

\begin{abstract}
We present results of an extensive search for molecular gas in 14 high
redshift ($z>2$) radio galaxies. Radio
galaxies with redshifts between 2.0 and 4.3 were observed 
during several sessions
on two different single-dish telescopes (IRAM and JCMT) and two interferometric
arrays (VLA and OVRO).
No significant CO emission 
to a 2$\sigma$ limit of a few times $10^{10}$ K\,km\,s$^{-1}$ pc$^2$
was detected in any of these objects.
The limits on the CO emission achieved indicate that, assuming a CO-H$_2$
conversion factor similar to the Galactic value, the mass of enriched
molecular gas 
in these galaxies is less than $10^{11}\,$M$_{\odot}$. This suggests
that distant radio galaxies may not be forming stars at the extremely high
rates that have earlier been proposed, although they may still be forming
stars at rates comparable to those of local starburst galaxies.
\keywords{Galaxies: active -- Galaxies: radio -- Galaxies:
molecular gas}
\end{abstract}

\section{Introduction}

One of the key questions in the evolution of galaxies
concerns the epoch of maximum star formation.
Since in the Milky Way star formation takes place in molecular clouds,
the presence of large amounts of molecular gas in distant galaxies would 
be an indication that they are undergoing prodigious star formation. For this 
reason it is important to find and study molecular gas in the most distant 
galaxies.

The first high redshift galaxy to be detected in molecular emission
lines was IRAS\,F10214+4724  at $z=2.286$ (\cite{row91}).
Several $\rm{CO}$ lines were detected (\cite{bro91,sol92a}) which would
indicate a
molecular gas mass of $\sim10^{11}\,$M$_{\odot}$\footnote{Throughout
this paper we assume a Hubble constant
$H_0=100$\,km\,s$^{-1}$ Mpc$^{-1}$ and a deceleration parameter
$q_0=0.5$; values quoted from other papers are,
where necessary, tacitly converted to this cosmology.}, dominating the
dynamical mass. However, recently it was discovered that 
IRAS$\,$F10214+4724 is
gravitationally lensed (\cite{bro95,gra95}).
Given the estimated lensing factor
between 3 and 20, its true molecular gas mass
may be only  $\sim10^{10}\,$M$_{\odot}$.
Since the discovery of CO emission from IRAS$\,$F10214+4724,
confirmed
molecular line emission has been detected in one other high redshift object,
the also gravitationally 
lensed ``Cloverleaf'' quasar H$\,$1413+117 at $z=2.5$ (\cite{bar94c}).
In spite of the uncertain lensing factor (estimated $\sim 7$)
this quasar must also contain a
large amount of molecular gas.
Noteworthy is that both the Cloverleaf quasar and IRAS$\,$F10214+4724 have
an AGN-type spectrum (see Elston {et al.}\ 1994)\nocite{els94}, 
i.e., a spectrum with high ionization lines
produced by a hard ultraviolet continuum. In spite of this
ionizing continuum there is a large mass in molecular gas apparently 
shielded from the UV radiation so that the molecules can survive.

In addition, there have been reports on possible CO emission from
damped Ly$\alpha$ absorption systems (\cite{bro93,fra94}), but more
sensitive observerations could not confirm these (Braine et al. 1996,
Wiklind and Combes 1994).

The discovery of the strong CO emission from IRAS$\,$F10214+4724
indicated that it might be possible to observe molecular gas out to the
highest currently known redshifts and suggested that many high redshift 
objects might contain large amounts of molecular gas.
High redshift radio galaxies (HZRGs) usually have very bright
Ly$\alpha$ emission, with integrated fluxes of 
$\sim10^{-15}\,$erg\, s$^{-1}$\, cm$^{-2}$
(luminosities $10^{44}$\,erg\,s$^{-1}$), and can extend over more than $10''$
($\sim100$\,kpc), indicating the presence of a large reservoir of gas in and
around these galaxies. The importance of HZRGs for the formation and evolution
of galaxies in general is shown by the fact that radio galaxies were almost a
thousand times more common at $z\sim2$ than at the current epoch. 
The bright 
emission lines in the optical spectra of HZRGs are a factor of $5-10$ 
narrower than those of
quasars, facilitating an accurate determination of their redshifts, 
which is a
prerequisite for deep mm-wave spectroscopy due to the limited
spectral bandwidth of current instrumentation.

It has been proposed that HZRGs are protogalaxies undergoing their
first major burst of star formation 
(see Eales and Rawlings 1993, and references
therein).
At low redshifts, radio galaxies have molecular gas masses of the
order of
10$^{10}\,$M$_{\odot}$ (\cite{maz93}). Such masses are comparable
to those of known luminous infrared starburst galaxies like 
Arp$\,$220 (\cite{sol90}).
High redshift radio galaxies have been suspected of having vigorous starbursts
associated with the propagation of the radio jet (\cite{ree89a,beg89}) as one
of the explanations of the alignment of the optical, infrared and radio axes
found in HZRGs (\cite{cha87,mcc87a}), although alternatives have been
proposed as well (cf., Sect.~4.2). 
The starburst model proposed by Chambers and Charlot (1990) \nocite{cha90a} 
requires the formation of 
$10^{11}\,$M$_{\odot}$ of stars over $10^7$ yr. 
Such starbursts should be accompanied by the production of large amounts of
heavy elements, giving rise to the formation of copious amounts of
dust and molecules. Evidence for the presence of dust comes from polarization
measurements of HZRGs (e.g., Tadhunter et al.\ 1992;
Cimatti {et al.}\ 1993)\nocite{cim93,tad92} 
and detections of
continuum emission from dust at submillimetre wavelengths (\cite{dun94,ivi95}).
The detection of large amounts of molecular gas in such objects would be
strong evidence in favour of the massive starburst hypotheses.
\nocite{eal93a}

In an extensive survey of ultrasteep spectrum radio sources we have
found 29 radio galaxies at $z>2$ (\cite{oji94a,oji96a,rot95a,rot96}),
a significant fraction of the total number of radio galaxies known at
such high redshifts.  The large luminosities of the observed emission
lines indicate that there is a powerful source of ionizing radiation
in such objects, albeit hidden from the observers.  Any molecular gas
would have to be shielded from this ionizing radiation (as in
IRAS$\,$F10214+4724 and the Cloverleaf quasar) to prevent the
molecules from dissociating.  The anisotropic ionizing radiation is
most likely oriented in a cone along the radio axis (\cite{ant93}).
Thus most molecular gas might be situated outside this cone of
ionizing radiation and is likely undisturbed by the radio jet, which
may be responsible for the large velocity dispersions of the extended
optical emission line gas (Van Ojik et al.\ 1996b). \nocite{oji96b}
Thus, we may expect any CO emission lines to be much narrower than the
optical emission lines (which have typically $1000\,$km\,s$^{-1}$
FWHM) and similar in width to the CO line widths observed both in
IRAS\,F10214+4724 and the Cloverleaf quasar ($\sim 300$\,km\,s$^{-1}$
FWHM).  We here report on our search for molecular gas in 14 HZRGs.

\begin{table}[t]
\caption[]{Single-dish observing sessions}
\begin{center}
\begin{tabular}{ccccl}  
\hline
\noalign{\smallskip}
Session & Date        &  Telescope & Time           & Comments \\
        &             &            & [hours]        & \\ 
\noalign{\smallskip}
\hline
\noalign{\smallskip}
1 & Nov 1993    & JCMT       & 56 & \\
2 & Nov 1993    & IRAM       & 20 & variable conditions \\
3 & Jan 1994    & IRAM       & 28 & \\
4 & Apr 1994    & JCMT       & 24 & \\
5 & Jun 1994    & JCMT       & 40 & variable conditions \\
6 & Aug 1994    & IRAM       & 103 & variable conditions \\
7 & Dec 1994    & JCMT       & 88 & \\ 
8 & Apr 1995    & IRAM       & 55 & variable conditions \\ 
\noalign{\smallskip}
\hline
\end{tabular}
\end{center}
\end{table}

\section{Observations and reduction}

\subsection{Transitions surveyed}

Our targets have redshifts ranging from 2.0 to 4.3. Taking into account the
instrumentation available, this allowed us to survey CO transitions from
$J=3{\to}2$ (345.8\,GHz rest frequency) for the objects with $z\sim2$
observed at IRAM or the Owens Valley Radio Observatory millimeter 
interferometer (OVRO)
in the 3\,mm window, to $J=9{\to}8$ (1036.9\,GHz rest frequency) for the
objects with the highest redshifts observing at the JCMT in the
1.3\,mm window.
With the VLA
the CO $J=1{\to}0$ transition can be observed at 25\,GHz 
for the sources with the 
highest redshifts.

Strategies for detecting CO at high $z$
have been discussed by Van der Werf and Israel (1996), who show that
if conditions in these objects are similar to those in local starburst
galaxies, the $J{=}3{\to}2$ to $J{=}6{\to}5$ lines are the transitions of
choice in detection experiments. Significantly, the line ratios found in
IRAS$\,$F10214+4724 and the Cloverleaf quasar are indeed similar to
those in local starburst galaxies (\cite{sol92a,bar94c}).


With the exception of B2$\,$0902+34 (Lilly 1988)
and 8C\,1435+635 (Lacy et al.\ 1994), which were included because of
their thermal dust emission (\cite{chi94,ivi95}),
the sources observed were selected from the Leiden sample of HZRGs.
One target, 0211$-$122,
has an optical spectrum similar to that of IRAS$\,$F10214+4724, with 
strongly absorbed Ly$\alpha$ indicating the presence of dust (\cite{oji94a}). 
Four other targets were selected because they showed
evidence for associated H\,{\sc i} absorption 
(Van Ojik et al.\ 1996b; R\"ottgering et al.\ 1995a)
\nocite{oji96b,rot95a} and the remaining
targets were selected because of their large Ly$\alpha$ halos and high
redshifts.
Thus, the 14 objects we observed are representative of our general
sample of HZRGs.

\begin{table}[t]
\caption[]{Log of the single-dish CO observations, with the
r.m.s.~noise $\sigma$ in the spectrum given at a velocity resolution
of 100\,km\,s$^{-1}$.}
\begin{center}
\begin{tabular}{lcccccc} 
\hline
\noalign{\smallskip}
Source & $z$ & Session & Line & $t_{\rm int}$ & $\sigma$  \\
       &           &   &         & [hours] & $T_{\rm mb}$
[mK] \\
\noalign{\smallskip}
\hline
\noalign{\smallskip}
0211$-$122 & 2.336 & 1      & CO(7--6)      &  25.0            & 2.0  \\
           &       & 2      & CO(3--2)      &   4.0            & 1.2  \\
           &       & 2      & CO(4--3)      &   3.6            & 1.3  \\
           &       & 6      & CO(3--2)      &   1.5            & 2.0  \\
           &       & 6      & CO(4--3)      &   1.5            & 2.0  \\
           &       & 6      & CO(7--6)      &   1.5            & 2.4  \\
           &       & 7      & CO(7--6)      &  14.3            & 1.2  \\
0214+183   & 2.131 & 7      & CO(6--5)      &   1.7            & 2.3  \\
0355$-$037 & 2.153 & 7      & CO(6--5)      &   3.0            & 2.3  \\
0448+091   & 2.040 & 7      & CO(6--5)      &   2.0            & 1.3  \\
4C\,41.17  & 3.800 & 2      & CO(4--3)      &   7.4            & 0.5  \\
           &       & 2      & CO(6--5)      &   7.6            & 0.4  \\
0748+134   & 2.410 & 6      & CO(3--2)      &   4.8            & 0.5  \\
           &       & 6      & CO(4--3)      &   4.8            & 3.0  \\
           &       & 6      & CO(7--6)      &   4.8            & 2.1  \\
           &       & 7      & CO(7--6)      &   2.3            & 1.9  \\
0828+193   & 2.572 & 3      & CO(3--2)      &   1.5            & 1.4  \\
           &       & 3      & CO(5--4)      &   1.5            & 1.7  \\
           &       & 6      & CO(3--2)      &   1.9            & 1.5  \\
           &       & 6      & CO(5--4)      &   1.9            & 3.8  \\
           &       & 7      & CO(7--6)      &   2.3            & 1.3  \\
B2\,0902+34& 3.395 & 2      & CO(4--3)      &   3.6            & 0.6  \\
           &       & 2      & CO(5--4)      &   5.6            & 0.8  \\
0943$-$242 & 2.918 & 4      & CO(8--7)      &  14.2            & 1.6  \\
1243+036   & 3.570 & 3      & CO(4--3)      &   3.9            & 1.2  \\
           &       & 3      & CO(6--5)      &   3.9            & 1.1  \\
           &       & 4      & CO(9--8)      &  14.2            & 1.0  \\
           &       & 5      & CO(9--8)      &  26.8            & 2.8  \\
           &       & 6      & CO(4--3)      &   6.1            & 2.8  \\
           &       & 6      & CO(6--5)      &   6.1            & 6.4  \\
           &       & 8      & CO(4--3)      &  35.1            & 0.6  \\
1435+635   & 4.257 & 6      & CO(4--3)      &   7.1            & 0.4  \\
           &       & 6      & CO(5--4)      &   7.1            & 2.2  \\
1707+105   & 2.345 & 3      & CO(3--2)      &   0.9            & 0.9  \\
           &       & 3      & CO(4--3)      &   0.9            & 1.3  \\
2202+128   & 2.704 & 7      & CO(8--7)      &   2.0            & 1.6  \\
2251$-$089 & 2.000 & 7      & CO(6--5)      &   1.9            & 2.2  \\
\noalign{\smallskip}
\hline
\noalign{\smallskip}
\end{tabular}
\end{center}
\end{table}

\subsection{Single-dish observations}

We observed the sources in 8 different observing sessions, summarized in
Tables~1 and 2.

Four sessions were carried out with the James Clerk Maxwell Telescope
(JCMT) on Mauna Kea, Hawaii. We used the SIS receiver A2 tuned so that
the redshifted frequency of the most suitable CO transition for each
object, at 220 to 250\,GHz, was in the upper sideband.  The telescope
half-power beam width (HPBW) at these frequencies is about $20''$.  As
backend we used the 2048 channel Dutch Autocorrelation Spectrometer
with a bandwidth of $750\,$MHz.  The single sideband (SSB) system
temperatures in good atmospheric conditions were 300 to 450\,K\null.
We chopped the subreflector at $2\,$Hz with a throw of $\sim60''$ and
applied beam-switching.

Four sessions were carried out with the IRAM 30\,m telescope at Pico
Veleta, Spain. Usually, two or three receivers were used
simultaneously, so that up to three CO transitions could be observed
with high net observing efficiency.  Two 512 channel, $512\,$MHz wide
filter banks were used as backend for the 3\,mm and 2\,mm receivers
and an autocorrelator backend for the 1.3\,mm receiver. The HPBW was
$25''$ at 3\,mm, $17''$ at 2\,mm and $11''$ at 1.3\,mm.  The
subreflector was chopped at $0.5\,$Hz with a throw of $120''$.  System
temperatures at IRAM under reasonable weather conditions were
$200-400\,$K at $100\,$GHz, $300-500\,$K at $150\,$GHz and $\sim
1000$\,K at 230\,GHz (SSB).

The instantaneous
velocity coverage of a spectrum ranged from
$500\,$km\,s$^{-1}$ (JCMT at 230\,GHz) to
$1200$\,km\,s$^{-1}$ (IRAM 30\,m at
100\,GHz).
The CO lines from our HZRGs are expected to be faint and 
broad, probably of the order of 300\,km\,s$^{-1}$, covering a large
portion of an instantaneous spectrum. 
Thus, broad low level ripples in the
baseline of the spectra are a serious limitation in detecting these lines.  
In order to minimize baseline problems,
integrations were carried out at several different local oscillator (LO) 
settings. By extending the covered baseline, residual baseline
undulations should be better distinguishable and not be mistaken for
true CO emission.
Usually the LO settings were at 0, $+150$ and $-150\,$km$\,$s$^{-1}$ 
with respect to the calculated redshifted CO sky frequency, but occasionally
we extended the baseline to
600\,km\,s$^{-1}$ from the nominal sky frequency.
In this way we achieved a total velocity range of 
$\sim 1200-2000\,$km\,s$^{-1}$.

Reduction of the spectra from IRAM was done with
the CLASS (Continuum and Line Analysis Single-dish Software)
reduction package, while the JCMT spectra were processed with
the SPECX spectral line reduction package.
The observations of each observing shift and each velocity setting were
examined separately before summing them into a final spectrum. 
Scans with instrumental problems (e.g., curved baselines) or 
taken in poor conditions (atmospheric opacity 
$\tau > 0.2$) were discarded.
The final spectra were smoothed with a
$100$\,km\,s$^{-1}$ FWHM Gaussian 
(i.e., a factor of $2-3$ smaller than the expected 
width of any CO emission line), and resampled.
The antenna temperatures were converted to Rayleigh-Jeans main beam
brightness temperature outside the atmosphere ($T_{\rm mb}$ [K]).
The corresponding flux density scale for a point source with the JCMT is
18.9\,Jy\,K$^{-1}$ and with the IRAM telescope 4.7\,Jy\,K$^{-1}$ at 3
and 2\,mm and 4.8\,Jy\,K$^{-1}$ at 1.3\,mm.

\begin{table}[t]
\caption[]{Log of interferometric CO observations, with the
r.m.s.~noise $\sigma$ in the spectrum given at a velocity resolution
of 100\,km\,s$^{-1}$.}
\begin{center}
\begin{tabular}{lccccccc} \hline
Source & $z$ & Telescope & Line & $t_{\rm int}$ & $\sigma$  \\
       &          &                 &               &       [hrs]
& [mJy\,beam$^{-1}$] \\ 
\hline
\noalign{\smallskip}
1243+036 & 3.5856 & OVRO & 4--3 & 32  & 1.6 \\
1243+036 & 3.5856 & VLA  & 1--0 & 18  & 0.6 \\
1435+635 & 4.2576 & VLA  & 1--0 & 10  & 0.3 \\
\noalign{\smallskip}
\hline
\noalign{\smallskip}
\end{tabular}
\end{center}
\end{table}

\subsection{Interferometric observations}

Several systematic effects which limit deep spectroscopy using single-dish
radio telescopes are reduced or absent for interferometric
measurements, since interferometers only record correlated
signals and are thus not sensitive to artefacts arising in separate
elements. Two targets were therefore
observed with interferometric arrays to verify the reality of
apparent emission features
found during single-dish measurements.

The VLA was employed to search for CO emission from two of the highest 
redshift objects, 1243+036 ($z=3.6$) and 1435+635 ($z=4.3$). 
The VLA has the advantage of a large collecting area and 
sensitive receivers, but the disadvantage of a limited
bandwidth ($50\,$MHz maximum, or $650\,$km\,s$^{-1}$ at $z = 4$).
Moreover, limitations in the IF system at the VLA
constrain observations using a $50\,$MHz bandwidth to 
a discrete set of central frequencies, in sequential 
steps of 20\,MHz and 30\,MHz. For our observations we used
a series of overlapping 50\,MHz bands centered around the nominal systemic
redshift, with 8 spectral channels per band and two 
circular polarizations, synthesizing a total bandwidth of 130\,MHz for
1435+635 and of 80\,MHz for 1243+036. The observational cycle included 
a sequence of short (5 minute) scans at each frequency setting,
with scans on a nearby phase calibrator before and after each
source scan. 

The source 1435+635 was observed in the D-array (1\,km maximum baseline)
in May 1995.
The phase calibrator was 1435+638, and 3C\,286 was used for gain
and bandpass calibration. Four frequency settings were used implying
a total redshift range searched for emission of 4.2420 to 4.2732 
(observing frequencies around 21.9\,GHz).
The source 1243+036 was observed during reconfiguration from C array
(3\,km maximum baseline) to D array in October 1995. 
Only the D-array spacings were
employed in the final analysis in order to limit problems with phase
coherence. The source 3C\,273, which is located conveniently close to 1243+036 
at an angular distance of only 5$^{\circ}$, 
was used for phase and bandpass calibration,
and 3C\,286 was used for gain calibration. Three frequency settings
were employed, implying a redshift range searched for emission
of 3.5784 to 3.5932.
The observing frequencies were around 25.1\,GHz for this
source. This frequency is at the edge of the VLA K-band, leading to
LO phase lock problems on some antennas, and to a generally higher
noise level (by a factor 2.5 or so relative to the K-band center).
In the end, only 17 antennas were used in the analysis.
The beam (FWHM) of the VLA (in hybrid configuration BnC for 1243+036 and DnC
for 1435+635) was about $2.5''$ at 22\,GHz. 
This beam size is similar to the extent of optical continuum
emission.

Standard phase, amplitude, and bandpass calibration was employed
using the Astronomical Image Processing System (AIPS)
at the U.S.\ National Radio Astronomy Observatory, Socorro. 
Spectral channel images
were constructed for each source at each frequency setting,
and visually inspected for emission. Also, spectra were extracted 
at the position of the radio nucleus, and inspected for emission.
In all spectra the
theoretical noise level was achieved. One potential problem
with high frequency observations is phase coherence. 
To check the phase coherence, a continuum image  of each source
was synthesized. In both cases the continuum flux density
observed was close to the expected value (determined by extrapolating
a power law spectrum from lower frequencies to 22\,GHz), and the
spatial distribution agreed with that expected for each source.

The radio galaxy 1243+036 was also observed
at OVRO
during December 1994 and January 1995 at $100.5$\,GHz in redshifted 
CO(4--3).
Four tracks were obtained in various configurations of the six 
telescopes, with interferometer baselines from 35 to 242\,metres.
During every track 1243+036 was observed for a total time of about 8 
hours. Phases were calibrated using 3C\,273 
which was observed every 30 minutes.
Once every hour, measurements of 3C\,273 were also done
to correct the pointing. 3C\,273 was also used as a passband calibrator.
The flux density scale was calibrated with measurements of 3C\,84, 
3C\,454.3, 0528+134, Neptune and Uranus. The backend was configured
with four adjacent and slightly overlapping bands, each covering 128\,MHz
in 32 channels. The CO(4--3) line from 1243+036 was searched for in
lower side band (LSB) 
at 100.5\,GHz. System temperatures were typically 200\,K (SSB).
After standard editing and calibration, AIPS 
was used to produce the final data cube at a spectral resolution of 
32\,MHz (96\,km\,s$^{-1}$).
The OVRO beam size at 100\,GHz
was about $2''$

The interferometric observations with the achieved noise levels 
are summarized in Table 3.

\begin{figure*}
\begin{center}
\mbox{\psfig{figure=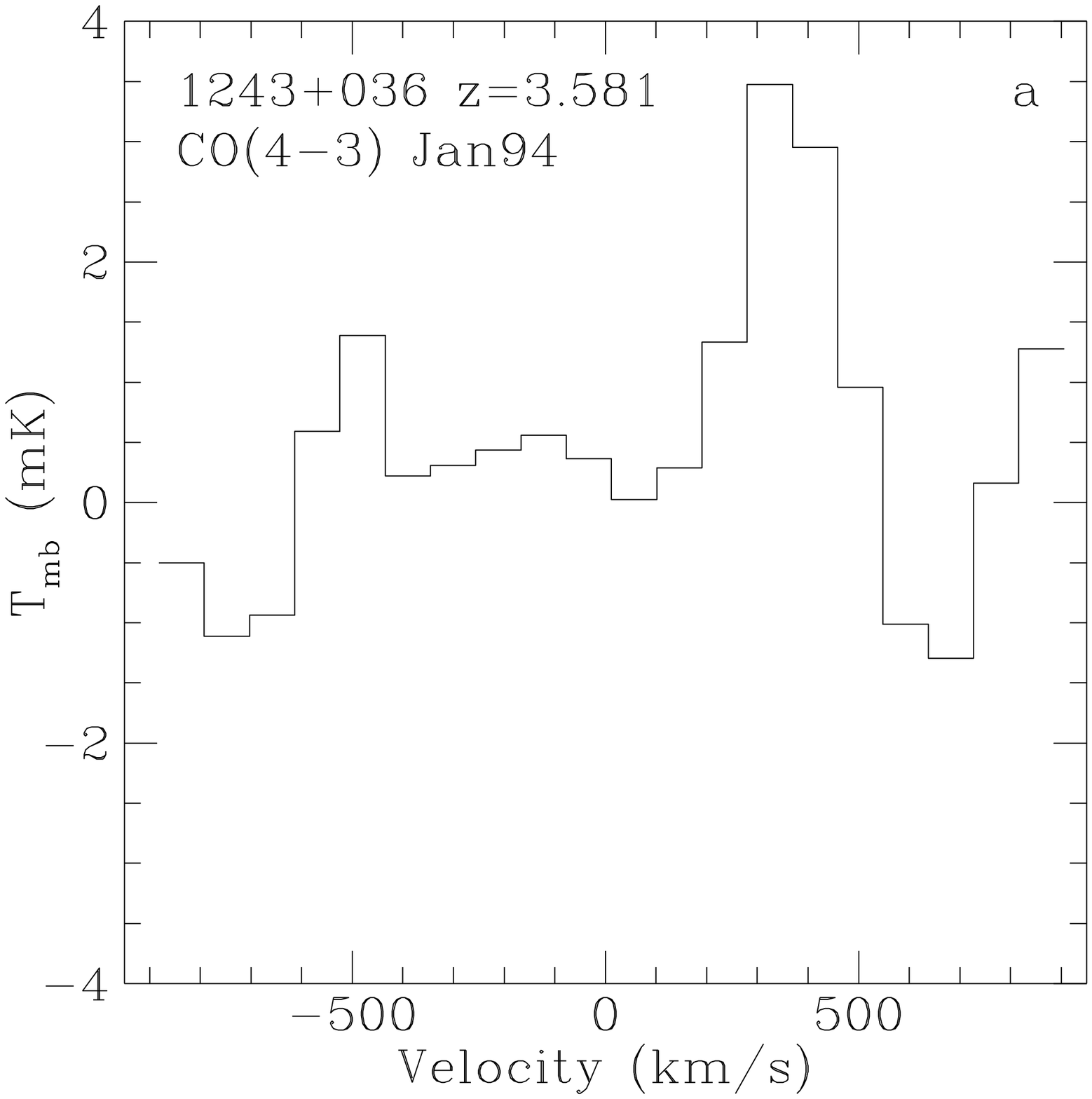,width=5.5cm,height=5.5cm}
\hfil
\psfig{figure=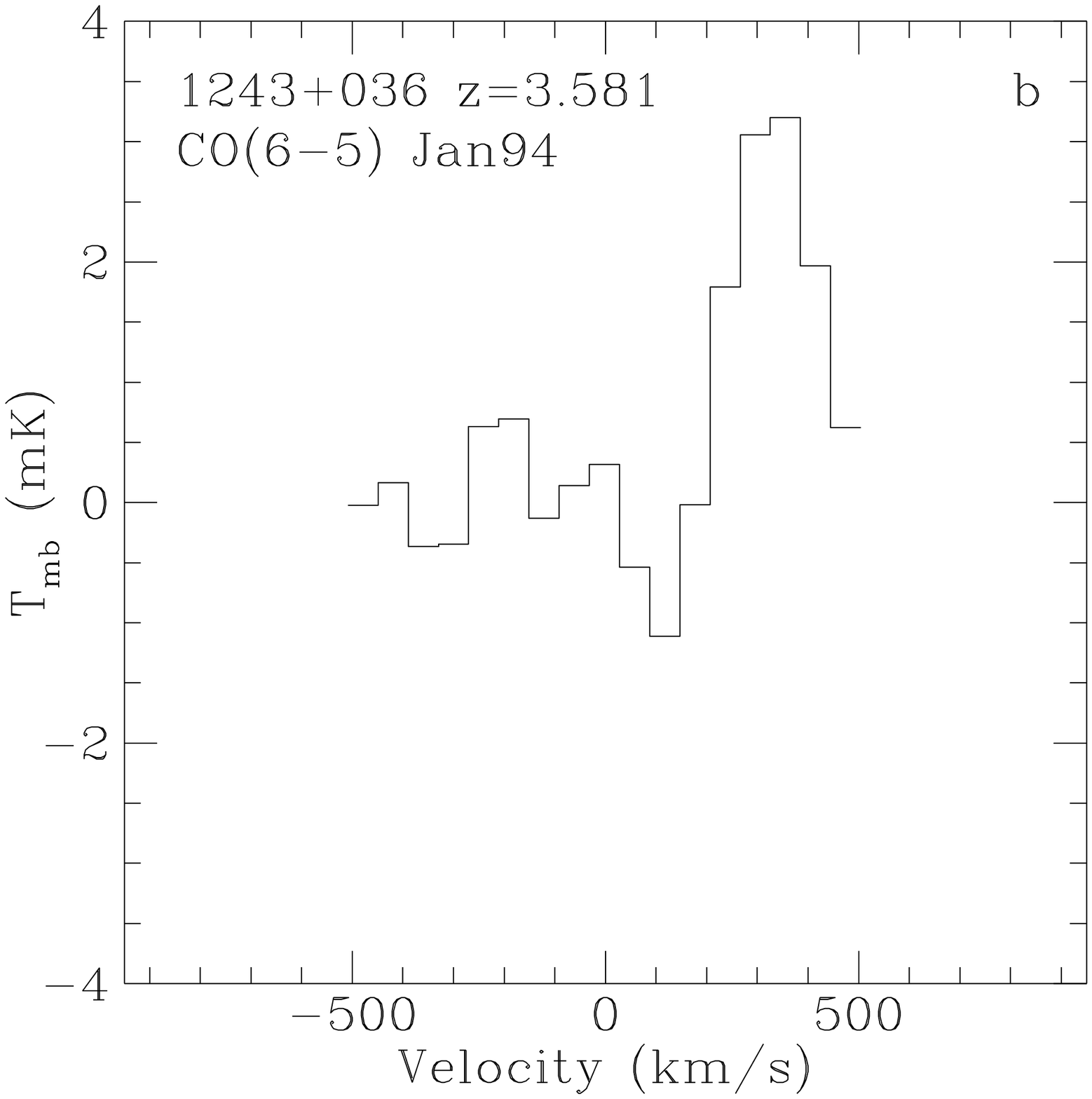,width=5.5cm,height=5.5cm}}
\mbox{\psfig{figure=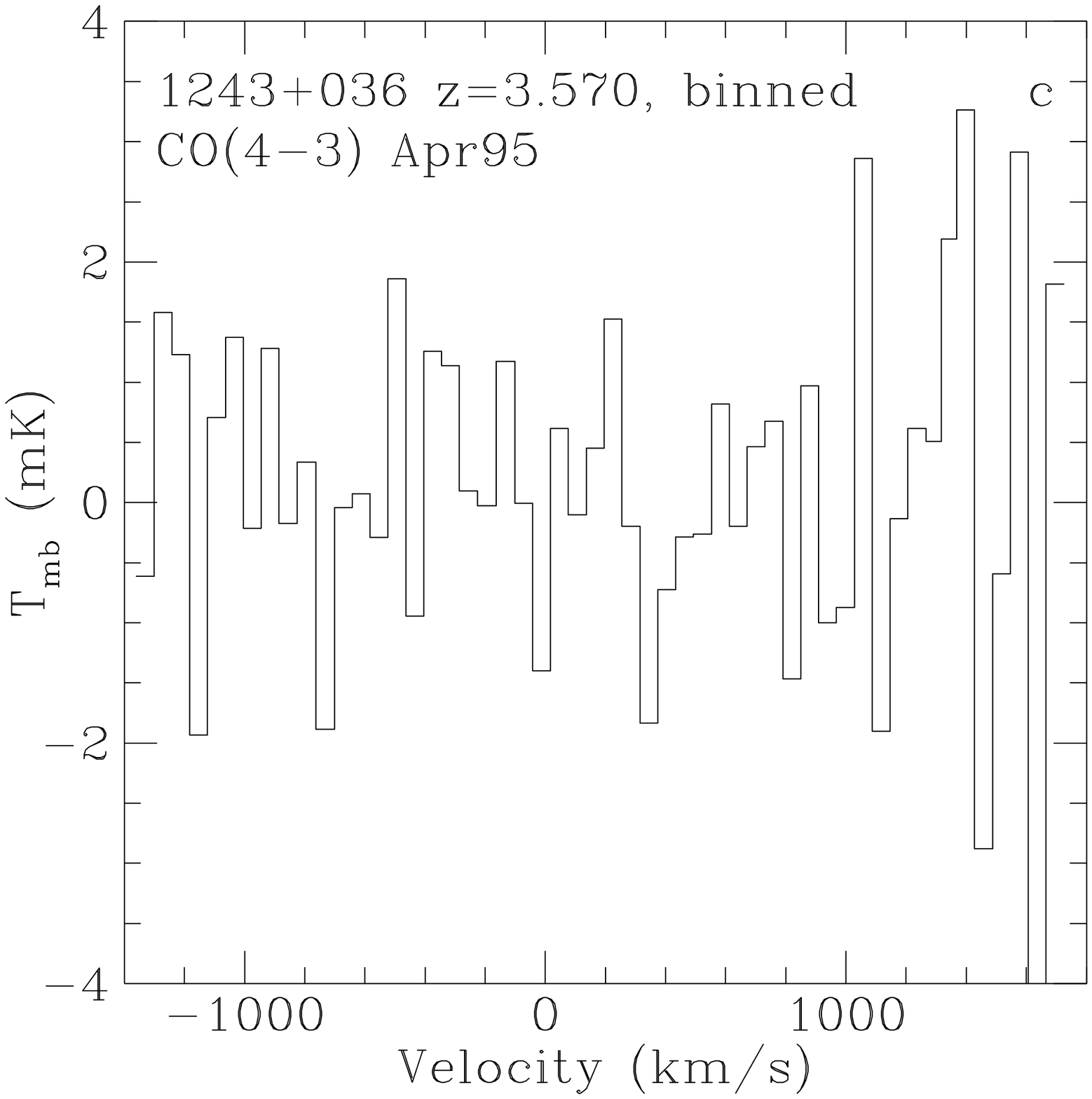,width=5.5cm,height=5.5cm}
\hfil
\psfig{figure=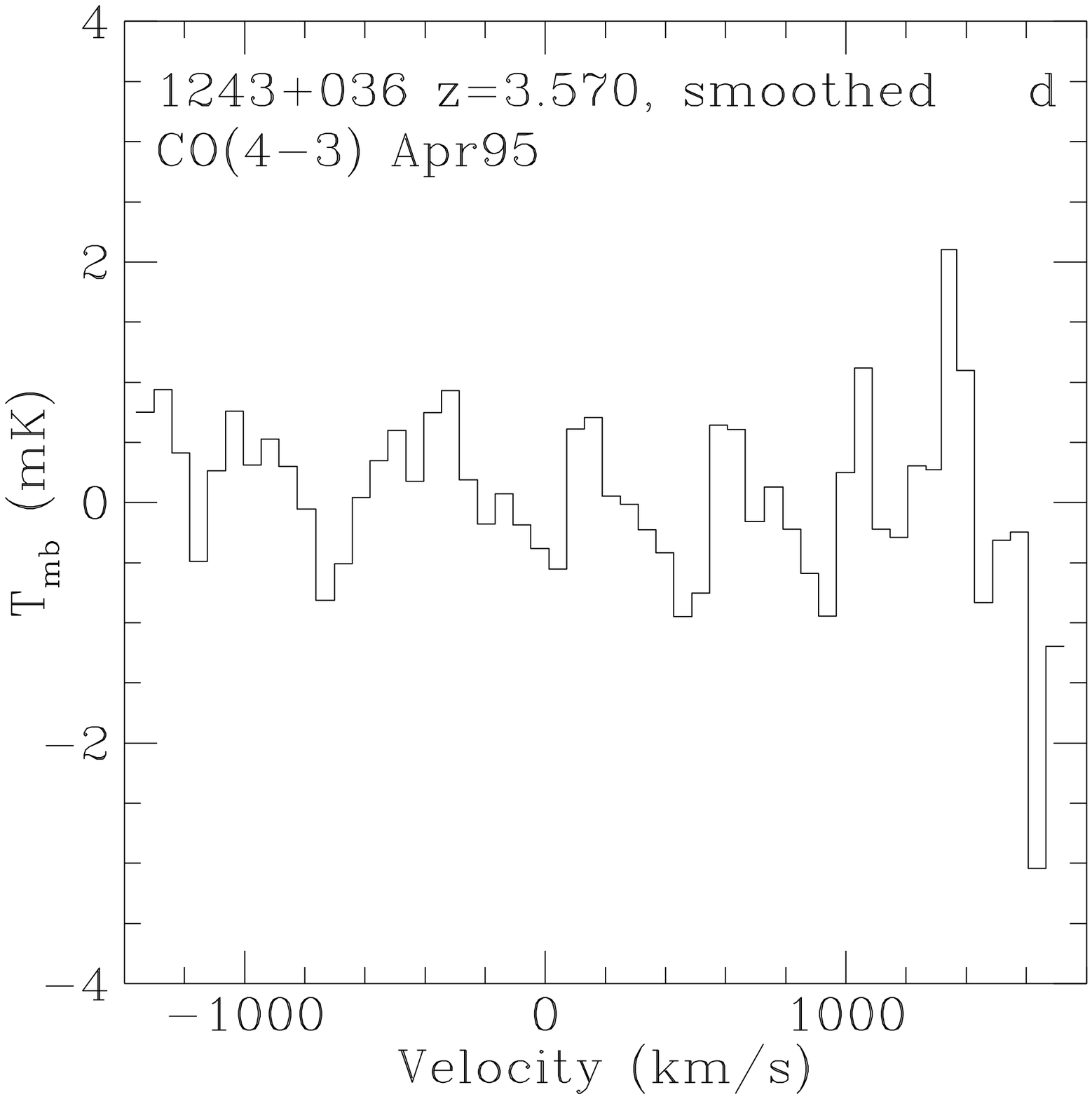,width=5.5cm,height=5.5cm}}
\end{center}
\caption[]{CO $J=4{\to}3$ (a) and $J=6{\to}5$ (b) spectra of the radio 
galaxy 1243+036 observed with the IRAM 30\,m telescope in January 1994. 
The spectra are centred at the redshift of Ly$\alpha$ from
a low resolution optical spectrum, were smoothed with a 
$100$\,km\,s$^{-1}$ FWHM Gaussian
and block-averaged 
into bins of $\sim 85$\,km\,s$^{-1}$.
Both in the 3\,mm {(a)} and the 2\,mm {(b)} 
spectrum a feature was observed at an offset velocity of 
$\sim 320$\,km\,s$^{-1}$ from the optical redshift. 
In spectra with different LO settings these features were
present at a constant sky frequency as expected for an
astronomical emission line.
The same object was observed in April 1995 with the IRAM 30m telescope
in CO $J=4{\to}3$ ((c) and (d)).
The centre is now at a
corrected redshift from a high resolution optical spectrum. We have extended
the baseline by adding many overlapping spectra with LO settings up to
$600\,$km\,s$^{-1}$ from the redshifted sky frequency.
In spectrum (c) the data was block-averaged into bins of
$65\,$km\,$^{-1}$. 
In spectrum (d), the data were first smoothed with a 
$100$\,km\,s$^{-1}$ FWHM Gaussian and then rebinned.
The feature seen in January 1994
should have appeared at $+1050\,$km\,s$^{-1}$. 
No evidence for any emission feature
is seen in (c) and (d), 
leading to the conclusion that the tentative result from the 
earlier observations (a) and (b) was spurious.}
\end{figure*}

\begin{figure*}
\begin{center}
\mbox{\psfig{figure=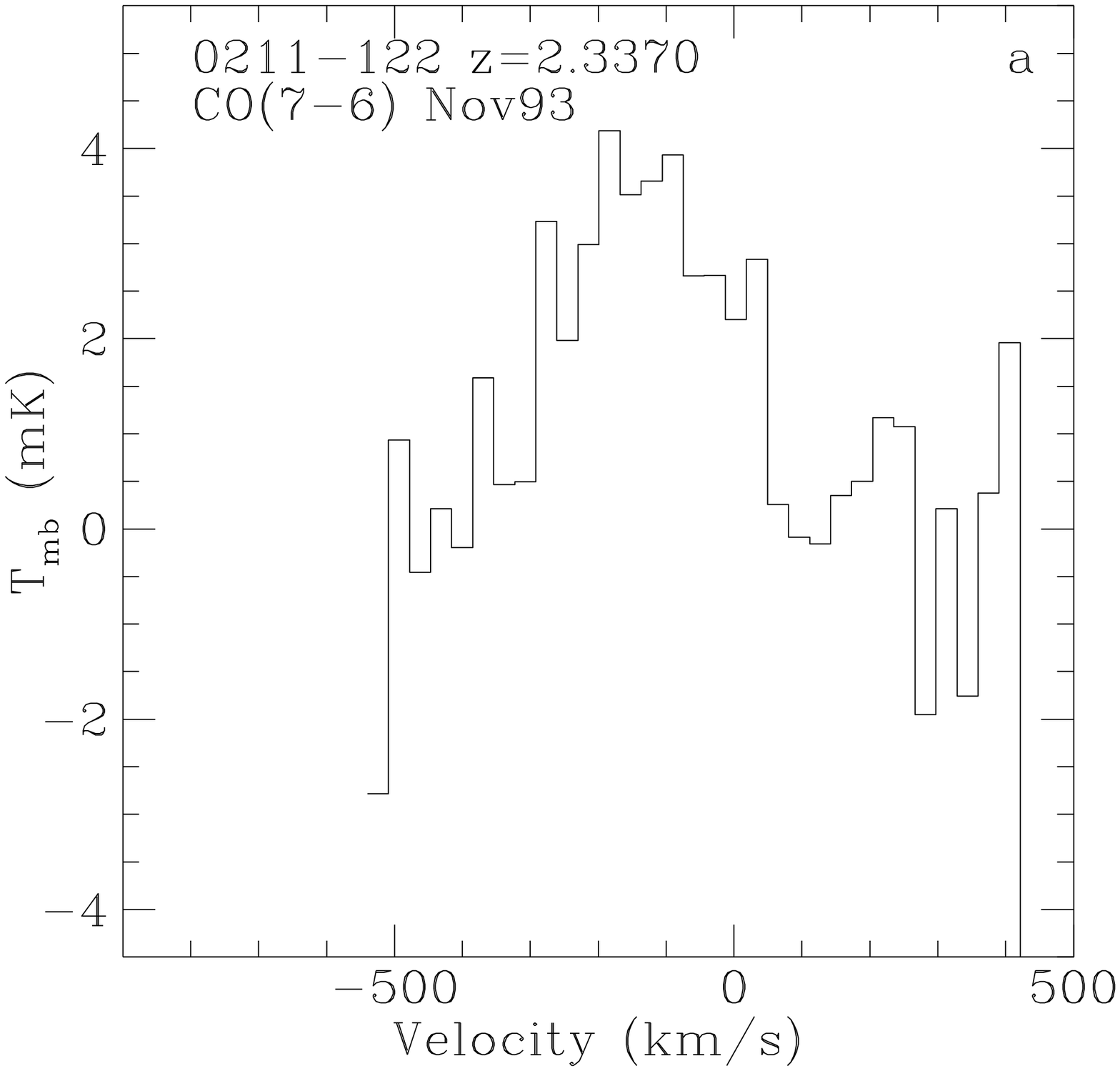,width=5.5cm,height=5.5cm}
\hfil
\psfig{figure=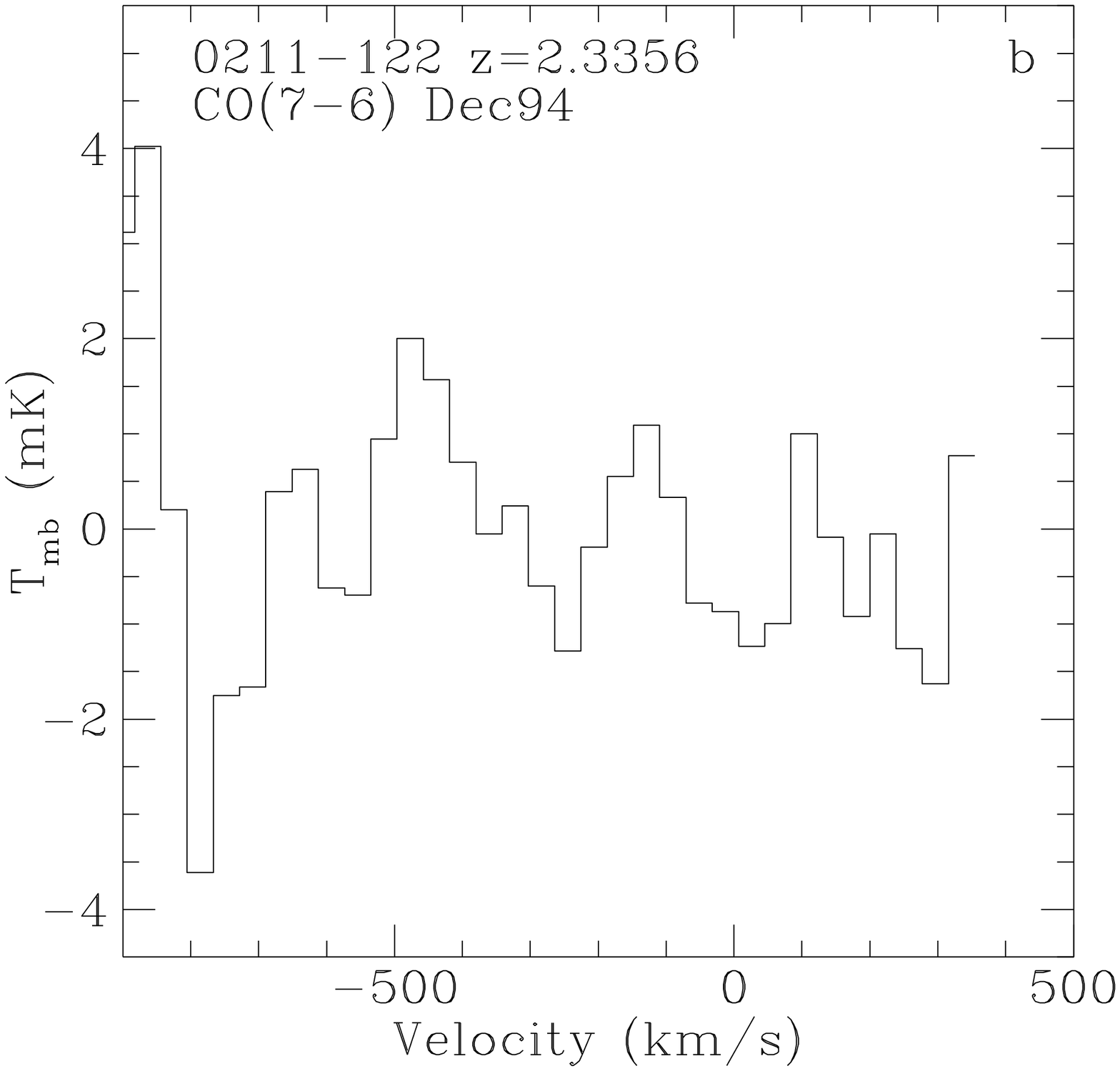,width=5.5cm,height=5.5cm}}
\end{center}
\caption[]{(a) Spectrum of CO(7--6) of the radio galaxy 0211$-$122
observed with the JCMT in November 1993. 
This spectrum includes only the data with 
atmospheric transparency $\tau < 0.1$, i.e. from 4 of the 6 shifts
(where one shift is 8 hours).
There appears to be a strong emission
feature centred at $-150$\,km\,s$^{-1}$ from the optical redshift.
(b) Spectrum of CO(7--6) of 0211$-$122 taken with the JCMT in December 1994,
centred at the tentative emission feature from spectrum (a). We have extended
the baseline by combining spectra with 5 different LO settings to better
distinguish any emission features.
The tentative emission feature is of (a) absent in (b).
Both spectra have been smoothed with a $100\,$km\,s$^{-1}$ FWHM
Gaussian and block-averaged into bins of 
$\sim60\,$km\,s$^{-1}$.}
\end{figure*}

\section{Results}

\subsection{Limits on CO emission from single dish observations}

A number of features were observed that significantly exceeded 
the noise level in the rest of the spectrum and remained when the
receivers were tuned to different offset velocities.
We first concluded that such features, at a level of a few mK,
were real CO emission features. 
However, none of these features 
could be reproduced in different observing sessions. When re-observed,
there was no evidence for any emission down to lower noise levels
or new features appeared at a different frequency. Examples of 
this situation are shown in Figs.~1 and 2.
Thus the observed features in the 
spectra were not due to line emission but were instrumental artefacts. 
The ability to detect faint broad emission lines depends strongly on these 
instrumental features in a spectrum and not only
on the formal noise level. A discussion of these effects in the JCMT
spectra has been given by R\"ottgering et al.\ (1995b).
%

\begin{table}[t]
\caption[]{$2\sigma$ upper limits on CO luminosity and molecular gas mass}
\begin{center}
\begin{tabular}{lcccc} \hline
Source & Line  & $z$ & line luminosity & $M$(H$_2$) \\
       &       &    & [K\,km\,s$^{-1}$ pc$^2$] &
[10$^{11}$\,M$_{\odot}$] \\ 
\hline
\noalign{\smallskip}
\multicolumn{5}{c}{Limits from the single dish observations} \\ 
\noalign{\smallskip}
\hline
\noalign{\smallskip}
0211$-$122 & CO(3--2)  & 2.336  & $< 2.4\times10^{10}$ & $< 1.0$ \\
0214+183   & CO(6--5)  & 2.131  & $< 4.1\times10^{10}$ & $< 1.6$ \\
0355$-$037 & CO(6--5)  & 2.153  & $< 4.1\times10^{10}$ & $< 1.7$ \\
0448+091 & CO(6--5)    & 2.040  & $< 2.1\times10^{10}$ & $< 0.8$ \\
4C\,41.17 & CO(4--3)   & 3.800  & $< 1.3\times10^{10}$ & $< 0.5$ \\
0748+134 & CO(3--2)    & 2.410  & $< 1.0\times10^{10}$ & $< 0.4$ \\
0828+193 & CO(3--2)    & 2.572  & $< 3.2\times10^{10}$ & $< 1.3$ \\
B2\,0902+34 & CO(4--3) & 3.395  & $< 1.3\times10^{10}$ & $< 0.5$ \\
0943$-$242 & CO(8--7)  & 2.918  & $< 2.6\times10^{10}$ & $< 1.0$ \\
1243+036 & CO(4--3)    & 3.570  & $< 1.3\times10^{10}$ & $< 0.5$ \\
1435+635 & CO(4--3)    & 4.257  & $< 1.1\times10^{10}$ & $< 0.5$ \\
1707+105 & CO(3--2)    & 2.345  & $< 1.8\times10^{10}$ & $< 0.7$ \\
2202+128 & CO(8--7)    & 2.704  & $< 2.3\times10^{10}$ & $< 0.9$ \\
2251$-$089 & CO(6--5)  & 2.000  & $< 3.5\times10^{10}$ & $< 1.4$ \\ 
\noalign{\smallskip}
\hline
\noalign{\smallskip}
\multicolumn{5}{c}{Limits from the interferometric observations} \\ 
\noalign{\smallskip}
\hline
\noalign{\smallskip}
1243+036 & CO(4--3)    & 3.570  & $< 1.0\times10^{10}$ & $< 0.4$ \\
1243+036 & CO(1--0)    & 3.570  & $< 6.5\times10^{10}$ & $< 3.3$ \\
1435+635 & CO(1--0)    & 4.257  & $< 2.8\times10^{10}$ & $< 1.3$ \\ 
\noalign{\smallskip}
\hline
\noalign{\smallskip}
\end{tabular}
\end{center}
\end{table}

We determined the final r.m.s.\ noise level of the spectra at a velocity 
resolution of 100\,km\,s$^{-1}$ over the entire 
observed velocity range (thus also including any
possible spurious features) and these noise levels are listed in
Table~2. 
Because of the occurrence of spurious features, 
some spectra with relatively short integration times but no spurious features
have lower effective 
noise levels than other spectra with longer total integration times
but with spurious features.

From the noise levels we estimate upper limits for the line flux from
the objects. If we assume that a CO emission line has a Gaussian shape
with a specific FWHM, we estimate a 2$\sigma$ upper limit to
the CO flux as
$1.06\times{\rm FWHM} \times2\times\sigma_{100}$ where
$\sigma_{100}$ is the r.m.s\ noise in the coadded spectrum at
a velocity resolution of $100\,$km\,s$^{-1}$.
We assume a FWHM of 300\,km\,s$^{-1}$ as for the observed lines in 
IRAS$\,$F10214+4724 and the Cloverleaf quasar (\cite{sol92a,bar94c}).
From the line flux limits we 
derive a 2$\sigma$ upper limit on the CO luminosity based on the lowest
CO transition observed under good atmospheric conditions, using 
the following formula, which is independent of beam size
(see Solomon {et al.}\ 1992a): 
\begin{equation}
L'_{\rm CO} = 3.25 \times 10^7\,S_{\rm CO}\Delta v\,
\nu_{\rm obs}^{-2}\,D_{\rm L}^2\,(1+z)^{-3},
\end{equation}
where $S_{\rm CO}\Delta v$ is the integrated line flux limit
in Jy\,km\,s$^{-1}$,
$\nu_{\rm obs}$ is the observing frequency in GHz, $D_{\rm L}$ 
the luminosity distance
to the object in Mpc and the line luminosity $L'$ is expressed in
K\,km\,s$^{-1}$\,pc$^2$.
The limits derived for the CO transitions used are listed in Table~4.

\subsection{Limits on CO and dust emission from interferometric observations}

No emission was seen in any spectral channel for either of the
two objects observed with the VLA and OVRO\null.
We derived $2\sigma$ upper limits
for the CO luminosity for the interferometric observations in the same
way as for the single dish observations. The limits are listed in Table~4.

The VLA continuum maps derived from the observations of 
1243+036 (25 GHz) and 1435+635 (22 GHz) do show continuum emission.
In 1243+036 there is a continuum peak of $3.5\,$mJy beam$^{-1}$ at the location
of the brightest hot spot seen in 8 GHz radio imaging observations 
(\cite{oji96a}).
The continuum emission of 1435+635 has a peak of 2.5\,mJy beam$^{-1}$ and shows
structure similar to the 8 GHz radio image.
The continuum flux densities are consistent
with non-thermal emission from the radio sources extrapolated using the
spectral index determined between 
8\,GHz to 22\,GHz. Thermal dust emission, as was recently
detected from 1435+635 (\cite{ivi95}), would
at a rest frequency of 115\,GHz only 
be expected at the $\mu$Jy level (estimated from the
spectrum of Ivison 1995). The radio continuum is at the level of 
$\sim 3$\,mJy at this frequency. Thus any emission from dust 
is negligible compared to the synchrotron radio emission.

No continuum emission was observed in the 100\,GHz continuum map derived from
the OVRO observations of 1243+036 to a $2\sigma$ upper limit of 
$0.4\,$mJy\,beam$^{-1}$.

\section{Discussion}

\subsection{Limits on the molecular gas masses}

For Galactic giant molecular clouds 
the relation between CO luminosity and H$_2$ mass has been
determined as
\begin{equation}
\alpha=M({\rm H}_2)/L'_{\rm CO} =  2.2 n_{{\rm H}_2}^{1/2}/T_{\rm b} = 4.78
\end{equation}
(\cite{sol92a}), where T$_{\rm b}$ is the intrinsic line
brightness
temperature (which is constant for thermalized, opaque lines from 
different rotational levels), $n_{{\rm H}_2}$ is the H$_2$ density in
cm$^{-3}$,
$M({\rm H}_2)$ is the H$_2$ mass in M$_{\odot}$ and $L'_{\rm CO}$ has
been given in Eq.~(1).
The conversion factor $\alpha$ depends on the excitation conditions of
the molecular gas and may be either higher or 
lower in extremely luminous 
and active galaxies such as IRAS\,F10214+4724 and HZRGs than in the
Milky Way. More fundamentally, in order to be able to trace H$_2$ with CO
lines, the molecular gas must be enriched, and, while subsolar CO
abundances are allowed, the CO column densities must be
sufficient to produce optically thick $^{12}$CO lines.
Solomon {et al.} (1992a) show that for 
IRAS\,F10214+4724 the conversion 
factor $\alpha$ for the higher CO transitions
is similar to the Galactic value of 4.78 for CO $J=1{\to}0$.
Using their value $\alpha\approx4$
for the lowest 
CO transitions observed under good conditions, we derive upper limits
for the (enriched)
H$_2$ masses for our 14 HZRGs of slightly less than
$10^{11}\,$M$_{\odot}$. For consistency, we have used the same value for the 
CO-H$_2$ conversion for the limits from the interferometric observations 
of CO(1--0), because it differs only little from the Galactic value of 4.78.
The upper limits for the CO luminosity and the corresponding H$_2$ mass are 
listed in Table~4.

The limits on the H$_2$ masses that we find are marginally smaller 
than the molecular gas mass that was initially estimated for 
IRAS$\,$F10214+4724 ($10^{11}\,$M$_{\odot}$ 
before it was realized that it is lensed.
In low redshift radio galaxies,
Mazzarella {et al.}\ (1993)\nocite{maz93} found molecular gas 
masses of order 10$^{10}\,$M$_{\odot}$.
Thus, we find that radio galaxies at high redshifts are not much more 
gas-rich than those at low redshifts.
As a prototypical gas-rich starburst galaxy and ultra-luminous IRAS
galaxy (ULIRG) at low redshift,
Arp$\,$220 has a large molecular mass, 
$M({\rm H}_2) \sim 2 \times 10^{10}$\,M$_{\odot}$ (Solomon {et al.} 
1990\nocite{sol90} and references therein). Although it is possible that our
HZRGs contain amounts of molecular gas similar to the ULIRGs, they cannot
contain much more.

\subsection{Implications of these and other millimetre observations at high 
redshift}

The two high $z$
objects with confirmed detections of emission from molecular gas,
IRAS$\,$F10214+4724 and the cloverleaf quasar,
are both gravitationally lensed. This
could explain why the present and other 
searches for molecular gas in high redshift galaxies
and quasars have been unsuccessful so far and might be taken as a
discouragement for further searches. However, for the two objects detected 
the lensing amplification is
unlikely to be much more than a factor of 10--20 (\cite{bar94c}), which,
although 
rendering the true CO flux density too low for current instrumentation, may
be observable in the future. The amplification of the flux densities of
IRAS$\,$F10214+4724 
and the Cloverleaf quasar imply that the underlying systems 
are
not necessarily of extraordinary properties. Although still quite luminous and
with $\sim10^{10}$\,M$_{\odot}$ of molecular gas quite gas-rich galaxies, they
may well be ``normal'' active galaxies in the early universe, so that that
many active galaxies, quasars and radio galaxies at high $z$ may
contain similar amounts of molecular gas.

The detection of submillimetre continuum emission from a number of high
redshift radio galaxies and quasars in several searches
(\cite{mac94,dun94,isa94,ivi95}) has been used to estimate gas masses
using simple assumptions on the dust temperature and gas/dust ratio. 
The estimates of molecular gas masses of
$0.5-1\times10^{11}$\,M$_{\odot}$ from these continuum observations are not in
contradiction with the non-detections of molecular gas in these objects and
our sample of HZRGs. Far-infrared luminosities of 
$10^{13}-10^{14}$\,L$_{\odot}$, as implied by the submillimetre dust 
detections, imply star
formation rates of $10^3-10^4$\,M$_{\odot}$\,yr$^{-1}$ if the FIR luminosity
is powered by star formation. By extrapolating from the somewhat lower
luminosity ($L_{\rm FIR}> 10^{12}$\,L$_{\odot}$) ultra-luminous IRAS galaxies
(\cite{san86}), a value $L_{\rm FIR}/M_{{\rm H}_2}$ of 
$100-1000\,$L$_{\odot}$/M$_{\odot}$ 
is expected at these
luminosities. The implied molecular masses are of the order of 
$10^{11}\,$M$_{\odot}$, only slightly more than the present upper limits.

One of the explanations proposed for
the strong alignment of the optical, near-infrared and
radio axes of HZRGs (\cite{cha87,mcc87a}) 
is that vigorous starbursts are
associated with the propagation of the radio jet (\cite{ree89a,beg89}).
In some of 
these radio jet-induced star formation scenarios, star formation rates of
the order of $10^4$\,M$_{\odot}$\,yr$^{-1}$ over a period of $\sim
10^8$\,yrs are
invoked to explain the alignment effect (\cite{cha90a,cha90}). 
Such star formation rates would
imply that a mass of order $10^{12}$\,M$_{\odot}$ in 
(molecular) gas is
converted into stars. The fact that we do not observe such quantities of
(enriched) molecular gas in any high redshift radio galaxy is relevant to this
scenario
and can have two explanations:
\begin{enumerate}
\item
The HZRGs that we observe have already consumed most of their molecular
gas and formed the bulk of their stellar population. 
Jet-induced star formation may have played a role, but we are
observing the object in a
later phase where the radio source has already expanded into a powerful
Faranoff-Riley~II (FR-II)
type source with an aligned stellar component in the galaxy hosting the radio
source. Mazzarella {et al.} (1993)\nocite{maz93} found molecular gas in
several low redshift radio galaxies detected with IRAS, but not in the most
powerful FR-II radio galaxies. They suggested that the powerful FR-II galaxies
might be in a later phase where most of the molecular gas has been
converted into stars.
\item
The alignment effect may not be due to jet-induced star
formation, but to another mechanism (cf., McCarthy 1993 for a review) 
such as scattering of 
continuum radiation emitted
along the radio axis from a hidden AGN (\cite{tad92,cim93,ser94}),
anisotropic
density distributions in the surrounding gas (Eales 1992) and inverse
Compton scattering (Daly 1992). 
The passage of the radio
jet still enhances star formation, but star formation would be not much more
than in local starburst galaxies, up to $\sim
100\,$M$_{\odot}$\,yr$^{-1}$.
\end{enumerate}
The possibility of a large amount of very cold molecular mass 
(temperature of a few K, 
as in the inner disk of M31, Allen \& Lequeux 1993)
\nocite{all93}
is ruled out, because the gas cannot be colder than
the cosmic microwave background which is $10\,$K at $z
\sim 2.5$, 
so that CO lines up to $J=3{\to}2$ 
would always be relatively strong (\cite{sol92b}).
Although such cold gas may show no emission from the higher
CO transitions, the lack of CO(1--0) and CO(3--2) 
emission observed in our data
indicates that even if the CO is cold, there must be less than
$\sim 10^{11}$\,M$_{\odot}$ in molecular gas. 

It is unlikely that metallicity effects play a major role. The CO-H$_2$
conversion factor $\alpha$ scales with CO abundance $X({\rm CO})$ 
approximately as $\alpha\propto [X({\rm CO})]^{-1/4}$ (Radford et al.\
1991; Solomon et al.\
1992a) as long as the CO lines are optically thick. Abundance effects
thus only begin to play a role when the CO
lines become optically thin, which requires an abundance of several
orders of magnitude below solar. While this situation cannot be ruled
out, the presence of large masses of dust as revealed by thermal
submm emission in several of our targets, and the
high metallicities indicated
by the optical spectrum of 0211$-$122 (Van Ojik et al.\ 1994) indicate
otherwise. Simultaneously, the dust will shield the molecules from
dissociation by the UV continuum, as is also the case in
IRAS$\,$F10214+4724 and the Cloverleaf quasar (cf., Sect.~1).
\nocite{mcc93}

\section{Summary and Conclusions}

We have searched for CO emission in a sample of 14 high redshift radio 
galaxies between $z=2$ and 4.3. In none of these galaxies have we found
consistent evidence for CO emission.

For a CO-H$_2$ conversion factor similar to the Galactic value, we find
2$\sigma$ upper limits to the molecular mass of less than 
$10^{11}$\,M$_{\odot}$.
The upper limits to the molecular gas mass in HZRGs suggests that they are 
not forming stars at rates of $\sim 10^4\,$M$_{\odot}$ yr$^{-1}$ which was
previously suggested in some jet induced star formation scenarios to explain the
optical-radio alignment effect.
This may mean that (i) all powerful HZRGs have already converted their gas
into stars, possibly through jet induced star formation, or 
(ii) jet induced star formation
does not play an important role in the alignment effect but HZRGs
may still be star forming at the rate of local starburst galaxies.


\begin{acknowledgements}These observations would not have been possible
without the good support of the staff at the IRAM 30\,m telescope 
and the James Clerk Maxwell Telescope. 
We thank Nick Scoville and Min Yun for carrying out the OVRO
observations and for useful discussions.
The James Clerk Maxwell Telescope is operated by The Observatories on 
behalf of the Particle Physics
and Astronomy Research Council of the United Kingdom, the Netherlands 
Organization for Scientific
Research, and the National Research Council of Canada.
OVRO is operated under funding from the U.S.
National Science Foundation, grant 93-14079.
The VLA is a facility of the U.S. National Radio
Astronomy Observatory, which is operated by Associated Universities, Inc.,
under cooperative agreement with the U.S. National Science Foundation.
We acknowledge support from an EC twinning project and a programme subsidy
granted by the Netherlands Organization for Scientific Research (NWO).
The research of Van der Werf has been made possible by the
Royal Netherlands Academy of Arts and Sciences.
\end{acknowledgements}


\begin{thebibliography}{}

\bibitem[{Allen \& Lequeux} {1993}]{all93}
Allen, R.J., Lequeux, J., 1993,
\newblock {ApJ} {410}, L15

\bibitem[{Antonucci} {1993}]{ant93}
Antonucci, R., 1993,
\newblock {ARA\&A} {31}, 473

\bibitem[{Barvainis {et~al.}} {1994}]{bar94c}
Barvainis, R., Tacconi, L., Antonucci, T., et al., 1994,
\newblock {Nat} {371}, 586

\bibitem[{Begelman \& Cioffi} {1989}]{beg89}
Begelman, M.C., Cioffi, D.F., 1989,
\newblock {ApJ} {345}, L21

\bibitem[\protect\citename{{Braine} et~al.}{1996}]{bra96}
{Braine}~J.,  {Downes}~D.,    {Guilloteau}~S.,  1996,
\newblock { ApJ}, { 309}, L43

\bibitem[{Broadhurst \& Leh\'ar} {1995}]{bro95}
Broadhurst, T., Leh\'ar, J., 1995,
\newblock {ApJ}, {450}, L41

\bibitem[{Brown \& {Vanden Bout}} {1991}]{bro91}
Brown, R.L., {Vanden Bout}, P.A., 1991,
\newblock {AJ} {102}, 1956

\bibitem[{Brown \& {Vanden Bout}} {1993}]{bro93}
Brown, R.L., {Vanden Bout}, P.A., 1993,
\newblock {ApJ} {412}, L21


\bibitem[{Chambers \& Charlot} {1990}]{cha90a}
Chambers, K.C., Charlot, S., 1990,
\newblock {ApJ} {348}, L1

\bibitem[{Chambers {et~al.}} {1987}]{cha87}
Chambers, K.C., Miley, G.K., Van Breugel, W., 1987,
\newblock {Nat} {329}, 604

\bibitem[{Chambers {et~al.}} {1990}]{cha90}
Chambers, K.C., Miley, G.K., Van Breugel, W.J.M., 1990,
\newblock {ApJ} {363}, 21

\bibitem[{Chini \& Kr\"ugel} {1994}]{chi94}
Chini, R., Kr\"ugel, E., 1994,
\newblock {A\&A} {288}, L33

\bibitem[{Cimatti {et~al.}} {1993}]{cim93}
Cimatti, A., Di Serego Alighieri, S., Fosbury, R.A.E., et al.\ 1993,
\newblock {MNRAS} {264}, 421

\bibitem[{Daly} {1992}]{}
Daly, R., 1992,
\newblock {ApJ} {399}, 426

\bibitem[{Di Serego Alighieri {et~al.}} {1994}]{ser94}
Di Serego Alighieri, S., Cimatti, A., Fosbury, R.A.E., 1994,
\newblock {ApJ} {431}, 123

\bibitem[{Dunlop {et~al.}} {1994}]{dun94}
Dunlop, J.S., Hughes, D.H., Rawlings, S., Eales, S.A., Ward, M.J.,
  1994,
\newblock {Nat} {370}, 347

\bibitem[{Eales} {1992}]{}
Eales, S.A., 1992,
\newblock {ApJ} {397}, 49

\bibitem[{Eales \& Rawlings} {1993}]{eal93a}
Eales, S.A., Rawlings, S., 1993,
\newblock {ApJ} {411}, 67

\bibitem[{Elston {et~al.}} {1994}]{els94}
Elston, R., McCarthy, P.J., Eisenhardt, P., et al., 1994,
\newblock {ApJ} {107}, 910

\bibitem[{Frayer {et~al.}} {1994}]{fra94}
Frayer, D.T., Brown, R.L., \& Vanden Bout, P.A., 1994,
\newblock {ApJ} {433}, L5

\bibitem[{Graham \& Liu} {1995}]{gra95}
Graham, J.R., Liu M.C., 1995,
\newblock {ApJ}, {449}, L29

\bibitem[{Isaak {et~al.}} {1994}]{isa94}
Isaak, K.G., McMahon R.G., Hills, R., Withington, S., 1994,
\newblock {MNRAS} {263}, L28

\bibitem[{Ivison} {1995}]{ivi95}
Ivison, R.J., 1995,
\newblock {MNRAS}, 275, L33

\bibitem[{Lacy {et~al.}} {1994}]{lac94}
Lacy, M., Miley, G.K., Rawlings, S., et al., 1994,
\newblock {MNRAS} {271}, 504

\bibitem[{Lilly} {1988}]{lil88}
Lilly, S., 1988,
\newblock {ApJ} {333}, 161

\bibitem[{McMahon {et~al.}} {1994}]{mac94}
McMahon, R.G., Omont, A., Bergeron, J., et al., 1994,
\newblock {MNRAS} {267}, L9


\bibitem[{Mazzarella {et~al.}} {1993}]{maz93}
Mazzarella, J.M., Graham, J.R., Sanders, D.B., Djorgovski, S., 1993,
\newblock {ApJ} {409}, 370

\bibitem[{McCarthy {et~al.}} {1987}]{mcc87a}
McCarthy, P.J., Van Breugel, W., Spinrad, H., Djorgovski, S., 1987,
\newblock {ApJ} {321}, L29

\bibitem[{McCarthy} {1993}]{mcc93}
McCarthy, P.J., 1993,
\newblock {ARA\&A} {31}, 639

\bibitem[{Radford et~al.} {1991}]{}
Radford, S.J.E., Solomon, P.M., Downes, D., 1991,
\newblock {ApJ} {369}, L15

\bibitem[{Rees} {1989}]{ree89a}
Rees, M.J., 1989,
\newblock {MNRAS} {239}, 1P

\bibitem[{R{\"o}ttgering et~al.}{1995a}]{rot95a}
R{\"o}ttgering~H.,  Hunstead~R.,  Miley~G.~K.,  van Ojik~R.,    Wieringa~M.~H.,
   1995a,
\newblock { MNRAS}, { 277}, 389

\bibitem[{R{\"o}ttgering et~al.}{1995b}]{rot95b}
R{\"o}ttgering~H.,  Jenness~T.,  Sleath~J.,  Miley~G.,  {van Ojik}~R.,    {van
  der Werf}~P.,  1995b,
\newblock { JCMT Newsletter}, { March}, 32

\bibitem[{R{\"o}ttgering et~al.}{1996}]{rot96}
R{\"o}ttgering~H.,  van Ojik~R.,  Miley~G.,  Chambers~K.,  van Breugel~W.,
  de~Koff~S.,  1996,
\newblock A \& A: in press

\bibitem[{Rowan-Robinson {et~al.}} {1991}]{row91}
Rowan-Robinson, M., Broadhurst, T., Lawrence, A., et al., 1991,
\newblock {Nat} {351}, 719

\bibitem[{Sanders {et~al.}} {1986}]{san86}
Sanders, D.B., Scoville, N.Z., Young J.S., et al., 1986,
\newblock {ApJ} {305}, L45

\bibitem[{Solomon {et~al.}} {1990}]{sol90}
Solomon, P.M., Radford, S.J.E., Downes, D., 1990,
\newblock {ApJ} {348}, L53

\bibitem[{Solomon {et~al.}} {1992a}]{sol92a}
Solomon, P.M., Downes, D., Radford, S.J.E., 1992a,
\newblock {ApJ} {398}, L29

\bibitem[{Solomon {et~al.}} {1992b}]{sol92b}
Solomon, P.M., Radford, S.J.E., Downes, D., 1992b,
\newblock {Nat} {356}, 318

\bibitem[{Tadhunter {et~al.}} {1992}]{tad92}
Tadhunter, C.N., Scarrott, S., Draper, P., Rolph, C., 1992,
\newblock {MNRAS} {256}, 53p

\bibitem[{Van der Werf \& Israel} {1996}]{}
Van der Werf, P.P., Israel, F.P., 1996,
\newblock In: Shaver, P. (ed.) Science with Large Millimetre
Arrays. Springer, in press
 
\bibitem[{van Ojik et~al.}{1994}]{oji94a}
van Ojik~R.,  R{\"o}ttgering~H.,  Miley~G.,  Bremer~M.,  Macchetto~F.,
  Chambers~K.,  1994,
\newblock { A\&A}, { 289}, 54

\bibitem[{van Ojik et~al.}{1996a}]{oji96a}
van Ojik~R.,  R{\"o}ttgering~H.,  Carilli~C.,  Miley~G.,    Bremer~M.,  1996a,
\newblock { A\&A}, { 313}, 25

\bibitem[{van Ojik et~al.}{1996b}]{oji96b}
van Ojik~R.,  R{\"o}ttgering~H. J.~A.,  Miley~G.~K.,    Hunstead~R.,  1996b,
\newblock A\&A: in press

\bibitem[{Wiklind \& Combes} {1994}]{wik94}
Wiklind, T., Combes, F., 1994,
\newblock {A\&A} {288}, L41

\end{thebibliography}
\end{document}